\documentclass[aps,twocolumn, prd, amsmath, floats,floatfix, superscriptaddress, nofootinbib, showkeys]{revtex4}
\usepackage{amssymb}
\usepackage{amsmath}
\usepackage{verbatim}
\usepackage{mathrsfs}
\usepackage{amsfonts}
\usepackage{latexsym}
\usepackage{epsfig}
\usepackage{color}
\usepackage{graphicx,subfigure}
\usepackage{units}
\usepackage{hyperref}
\usepackage{envmath}
\usepackage{natbib}
\usepackage{ctable}
\usepackage{multirow}
\usepackage{fancyhdr}
\usepackage{float}
\usepackage{hyperref}
\usepackage[version=3]{mhchem}
\makeatletter \newcommand*{\rom}[1]{\expandafter\@slowromancap\romannumeral #1@} \makeatother
\makeatother
\DeclareUnicodeCharacter{2212}{-}
\DeclareUnicodeCharacter{060C}{!!!FIX ME!!!}
\begin{document}
\definecolor{orange}{rgb}{0.9,0.45,0}
\def\CovDev{D}
\def\Res{{\mathcal R}}
\def\Gammaflat{\hat \Gamma}
\def\metricflat{\hat \gamma}
\def\Dflat{\hat {\mathcal D}}
\def\part_n{\partial_\perp}
%
\def\Lie{\mathcal{L}}
\def\A{\mathcal{X}}
\def\Aphi{\A_{\phi}}
\def\hAphi{\hat{\A}_{\phi}}
\def\E{\mathcal{E}}
\def\Ham{\mathcal{H}}
\def\M{\mathcal{M}}
\def\R{\mathcal{R}}
\def\p{\partial}
\def\hg{\hat{\gamma}}
\def\hA{\hat{A}}
\def\hD{\hat{D}}
\def\hE{\hat{E}}
\def\hR{\hat{R}}
\def\hcA{\hat{\mathcal{A}}}
\def\hDelt{\hat{\triangle}}
\def\na{\nabla}
\def\dif{{\rm{d}}}
\def\non{\nonumber}
\newcommand{\erf}{\textrm{erf}}
%
\renewcommand{\t}{\times}
\long\def\symbolfootnote[#1]#2{\begingroup%
\def\thefootnote{\fnsymbol{footnote}}\footnote[#1]{#2}\endgroup}
\title{A New Constraint on the Simulation of the Intergalactic Medium through the Evolution of the Neutral Hydrogen Fraction in the Epoch of Reionization}

\author{S. Mobina Hosseini}
\email{mobinahosseini954@gmail.com}
\affiliation{Department of Physics, K.N. Toosi University of Technology, P.O. Box 15875-4416, Tehran, Iran}
\affiliation{PDAT Laboratory, Department of Physics, K. N. Toosi University of Technology, P.O. Box 15875-4416, Tehran, Iran}

\author{Bahareh Soleimanpour Salmasi}
\email{physics.bahare@gmail.com}
\affiliation{Department of Physics, Shahid Beheshti University, P.O. Box 19839-69411, Tehran, Iran}
\affiliation{PDAT Laboratory, Department of Physics, K. N. Toosi University of Technology, P.O. Box 15875-4416, Tehran, Iran}

\author{Seyed Sajad Tabasi} 
\email{sstabasi98@gmail.com}
\affiliation{Department of Physics, Sharif University of Technology, P. O. Box 11155-9161, Tehran, Iran}
\affiliation{PDAT Laboratory, Department of Physics, K. N. Toosi University of Technology, P.O. Box 15875-4416, Tehran, Iran}

\author{Javad T. Firouzjaee}
\email{firouzjaee@kntu.ac.ir}
\affiliation{Department of Physics, K.N. Toosi University of Technology, P.O. Box 15875-4416, Tehran, Iran}
\affiliation{PDAT Laboratory, Department of Physics, K. N. Toosi University of Technology, P.O. Box 15875-4416, Tehran, Iran}
\affiliation{School of Physics, Institute for Research in Fundamental Sciences (IPM), P.O. Box 19395-5531, Tehran, Iran}

\begin{abstract}
The thermal history of the intergalactic medium is full of extremely useful data in the field of astrophysics and cosmology. In other words, by examining this environment in different redshifts, the effects of cosmology and astrophysics can be observed side by side. Therefore, simulation is our very powerful tool to reach a suitable model for the intergalactic medium, both in terms of cosmology and astrophysics. In this work, we have simulated the intergalactic medium with the help of the 21cmFAST code and compared the evolution of the neutral hydrogen fraction in different initial conditions. Considerable works arbitrarily determine many important effective parameters in the thermal history of the intergalactic medium without any constraints, and usually, there is a lot of flexibility for modeling. Nonetheless, in this work, by focusing on the evolution of the neutral hydrogen fraction in different models and comparing it with observational data, we have eliminated many models and introduced only limited simulation models that could confirm the observations with sufficient accuracy. This issue becomes thoroughly vital from the point that, in addition to restricting the models through the neutral hydrogen fraction, it can also impose restrictions on the parameters affecting its changes. However, we hope that in future works, by enhancing the observational data and increasing their accuracy, more compatible models with the history of the intergalactic medium can be achieved.
 
\end{abstract}

\maketitle

\section{Introduction}
 The physics of 21-cm radiation has recently significantly added to a better understanding of cosmology, particularly the nature of the Intergalactic Medium (IGM). The emission of the 21-cm line is dependent on several properties of the IGM, especially its density, neutral fraction, and spin temperature. Before the first Ultra-Violet (UV) sources become active, the spin temperature is determined by a competition between the tendency of radiative transitions to bring the spin temperature into equilibrium with the Cosmic Microwave Background (CMB) temperature, and the tendency of atomic collisions to bring the spin temperature into equilibrium with the gas kinetic temperature \cite{Hirata:2005mz}.

 It is crucial to take into account three wavelength regimes while thinking about first star radiation: the Wouthuysen-Field (WF) effect couples Lyman photons to the 21-cm line at high redshifts ($z\simeq 30$), X-ray photons from stellar remnants heat the gas, and Lyman-Werner (LW) photons (11.2-136 eV) breakdown molecular hydrogen, creating a negative feedback mechanism on star formation and decreasing the rate of heating 
\cite{Haiman:1996rc}. 
 
 During the formation of the first galaxies, UV radiation discharges into the IGM. Additionally, the first galaxies emit copious amounts of X-rays \cite{Mesinger:2012ys,Fialkov:2014kta}. This emission heats the IGM surrounding each galaxy, increasing the hydrogen spin temperature, and subsequently generating 21-cm emission, as the 21-cm temperature becomes positive for spin temperature greater than CMB temperature \cite{Munoz:2019hjh}. Changes to the background spectrum of photons can have some intriguing ramifications; for instance, exotic energy injection may explain observed excesses, such as the cosmic optical background \cite{Lauer:2022fgc,Bernal:2022wsu}.
 
 Plus, excessive Lyman-alpha $(Ly\alpha)$ can affect the global and inhomogenous redshifted 21-cm signal. As a function of redshift, the global signal is the mean 21-cm temperature across the entire sky. This radiation induced an excited state in the hydrogen atoms. The Lyman Coupling Era (LCE), which lasted approximately from the redshift 27 to 19, until the minimum of brightness temperature, when X-ray heating began to predominate, commenced with the formation of first stars and lasted until the onset of X-ray heating. This process is known as the WF effect, corresponding to the mixing of hyper-fine levels caused by $Ly\alpha$ photons \cite{wouthuysen1952,field1958}. What is more, background radiation has an impact on star formation; for example, photodetachment of intermediate states can disrupt the formation of molecular hydrogen which is essential in order to form the first stars \cite{Hirata:2006bt,Liu:2023nct}. 

 Another epoch is the Epoch of Heating (EoH), which begins at $z\simeq 19$ and continues until $z \simeq 14$ when the 21-cm signal crosses zero. In this epoch, the X-rays produced by the first galaxies heat the hydrogen gas and reduce the quantity of 21-cm absorption in an inhomogeneous manner. After the gas is completely heated, subsequent generations of stars begin reionizing the hydrogen by emitting UV light. This marks the beginning of the Epoch of Reionization (EoR), during which the global signal decays smoothly to zero as the fraction of neutral hydrogen (HI) gradually vanishes \cite{Furlanetto:2004nh,McQuinn:2005hk}.

 In the Milky Way, as well as nearby galaxies, radio observations of the 21-cm line are commonly used to map the velocity of the neutral hydrogen gas, but it has yet to be identified in emission at redshifts $z>1$. It is typical to utilize brightness temperature, to characterize the measured intensity of the 21-cm line when it operates as a cosmological probe. The background source is usually the CMB or a radio-bright point source, and the measured brightness temperature is compared to that. For cosmological purposes, it is preferable to consider the CMB, in which case the 21-cm signal appears as a spectral distortion across the entire sky \cite{Weltman:2018zrl}.

 Mapping the IGM fluctuations is exceedingly complicated due to the extremely weak spin-flip cosmological signal relative to other astrophysical radio backgrounds, and early attempts to detect it has concentrated on two complementary directions. Global signal measurement has been discovered in the literature very well \cite{Shaver:1999gb,Munoz:2020itp,Voytek:2013nua,Singh:2017syr,DiLullo:2020owx}. The Experiment to Detect the Global EoR Signature (EDGES) collaboration is the only one to have achieved a preliminary detection \cite{Bowman:2018yin}, even though the cosmological interpretation of the data is highly sensitive to instrumental and systematic uncertainties \cite{Sims:2019iyz,Tauscher:2020wso}. 
 
 Surprisingly, the claimed signal is much stronger than anticipated, demanding either that the IGM temperature falls lower than permitted by adiabatic cooling. This situation may be a manifestation of energy exchange with dark matter \cite{Barkana:2018lgd,Slatyer:2018aqg}.
 Upper limits have been reported from a variety of experiments from the redshift 6 to 10, such as Low-Frequency Array Radio-Telescope (LOFAR), though these limits have only marked the surface of the parameter space spanned by "standard" models of early galaxies
  \cite{Ghara:2020syx,Mondal:2020rce,Greig:2020hty}.
  However, other telescopes, such as the Hydrogen Epoch of Reionization Array (HERA) \cite{HERA:2021bsv,HERA:2021noe}, Giant Metrewave Radio Telescope (GMRT) \cite{Paciga:2010yy,Paciga:2013fj}, and Square Kilometre Array (SKA) \cite{Koopmans:2015sua}, will also work to gather cosmological and astrophysical data from the cosmic dawn and the EoR through the 21-cm signal.

  In SEC. \ref{SEC2} we have mentioned the theoretical equations around the 21-cm signals from the IGM. In SEC. \ref{SEC3} we have discussed the characteristics of the Epoch of Reionization. SEC. \ref{SEC4} expressed the Wouthuysen-Field effect in detail, then SEC. \ref{SEC5} clarifies the details of the 21cmFAST simulation and the initial conditions of this simulation. Ultimately, in SEC. \ref{SEC6} we have demonstrated models obtained by simulation and models compatible with observational data.
  
\section{General Equations\label{SEC2}}
 As it has been explained, the 21-cm wavelength is emitted from the spin-flip of the hydrogen atom.
The spin temperature is not a thermodynamical temperature, but a statistical temperature and indicates the ratio of the number density of the triplet state to the singlet state of hydrogen gas \cite{Furlanetto:2006tf}
\begin{equation}
    \label{e1}
    \frac{n_{1}}{n_{0}}=\frac{g_1}{g_0}\exp({-\frac{T_{*}}{T_{S}}}),
\end{equation}
 where $T_{*}=h\nu_{21}/k_B=0.0068K$  is the temperature that corresponds to hyper-fine transition. The fraction $g_1/g_0 = 3$ is the degeneracy level ratio of the triplet to the singlet state \cite{Widmark:2019cut}.
 There is a correlation between the hydrogen spin temperature, $T_{S}$, and the 21-cm difference in brightness temperature \cite{Pritchard:2011xb}
 \begin{equation}
 \begin{split}
     \label{e2} 
     \delta T_b & =27 x_{HI}(\frac{T_{S}-T_{\gamma}(z)}{T_{S}})(1+\frac{1}{H(z)}\frac{dv_{||}}{dr_{||}})^{-1}  \\
    & \times (\frac{1+z}{10})^{\frac{1}{2}}(\frac{\Omega_b}{0.044}\frac{h}{0.7})(\frac{\Omega_m}{0.27})^\frac{1}{2} mK,
  \end{split}
 \end{equation}
 \\
 where $x_{HI}$ is the local neutral fraction of hydrogen, which is determined by physical processes, particularly the recombination, the photoionization, and the collisional ionization, $\delta$ is the overdensity of the gas, $T_{\gamma}$ is the CMB temperature, $H(z)$ is the Hubble parameter, $dv_{||}/dr_{||}$ is the velocity gradient along the line of sight, $\Omega_m$ is the matter density, and $\Omega_b$ is the baryon density \cite{Semelin:2023lhz}.
 The ionization fraction also can be obtained as \cite{Peebles:1968ja}
\begin{align}
\begin{aligned}
\frac{dx_{e}}{dz} &= \frac{1}{H(z)(1+z)}\left[C_{p}\left(\alpha_{e}x_{e}^{2}n_{H}-\beta_{e}\left(1-x_{e}\right)e^{-\frac{h_{p}\nu_{\alpha}}{k_{B}T_{K}}}\right)\right. \\
&\left.-\gamma_{e}n_{H}\left(1-x_{e}\right)x_{e}\right],
\end{aligned}
\label{e3}
\end{align}
where $T_{K}$ is the kinetic temperature, $\beta_e$ is the photoionization coefficient, $C_p$ is the Peebles factor, $\alpha_e$ is the
recombination co-efficient, $h_p$ is the Planck constant, $\gamma_e$ is the collisional ionization
coefficient \cite{Minoda:2017iob}, and $n_{H}$ is the number density of hydrogen which is approximately $n_{H}\approx 0.189(1+z)^{3}$
   \cite{Cole:2019zhu}. It is noteworthy that $\beta_e$, $\alpha_e$, and $C_p$ has been explained in \cite{seager1999,Seager:1999km}. Plus, the ionizing photons and cosmic rays from the first formation of galaxies are expected to affect the ionization fraction \cite{Bera:2022kdk}.

 The 21-cm brightness temperature is also dependent on the optical depth, the spin temperature, and the CMB temperature \cite{Halder:2022ijw}
 \begin{equation}
 \label{e4}
     T_{21}=\frac{T_{S}-T_{\gamma}}{1+z}(1-\exp{(-\tau)}).
 \end{equation}
 Here, $\tau$ is the optical depth. For $\tau<<1$, the optical depth can be obtained
 \begin{equation}
 \label{e5}
     \tau=\frac{3c\lambda_{21}^{2}h A_{10}n_{H}}{32\pi k_{b}T_{S}H(z)},
 \end{equation}
 where $A_{10} \simeq 2.85\times10_{}^{-15}$ is the Einstein coefficient, $\lambda_{21}$ is the 21-cm wavelength, and $k_b$ is the Boltzmann constant.
 \\ The spin temperature is given by 
 \begin{equation}
 \label{e6}
 T_{S}^{-1}=\frac{T_{\gamma}^{-1}+x_{\alpha}T_{C}^{-1}+x_{c}T_{K}^{-1}}{1+x_{c}+x_{\alpha}},
 \end{equation}
 where $T_{C}$ is the color temperature of the $Ly\alpha$ radiation
field at the $Ly\alpha$ frequency, and $x_{c}$ and $x_{\alpha}$ are the coupling coefficients, then $x_{c}$ can be probed by following equation
\begin{equation}
\label{e7}
    x_{c}=\frac{T_{*}}{A_{10}T_{\gamma}}(\kappa_{10}^{HH}n_{H}+\kappa_{10}^{eH}n_{e}+\kappa_{10}^{pH}n_{p}),
\end{equation}
where hydrogen-hydrogen, electron-hydrogen, and proton-hydrogen collision de-excitation rates are denoted by $\kappa_{10}^{HH}$, $\kappa_{10}^{eH}$, and $\kappa_{10}^{pH}$ respectively \cite{Soltinsky:2021esh}. Using quantum mechanical considerations, these de-excitation rates are computed and fitted in different works ($\kappa_{10}^{HH}$ in \cite{Mittal:2021egv,Park:2005bu}, $\kappa_{10}^{eH}$ in \cite{Furlanetto:2006su}, and $\kappa_{10}^{pH}$ in \cite{Furlanetto:2007te}, over the range $1\le T_K \le 10^4K$).

Consequently, neutral hydrogen is significantly more abundant than free protons or electrons. It effectively makes up for its lower rate coefficient to spin exchange before the EoR. The ionized fraction of the IGM is approximately $10^{-4}$ before the EoR. As a result, at a relatively low level of the ionization percentage, hydrogen-hydrogen collisions account for the vast majority of the collisional contribution to the spin temperature. While the contributions of the other two types of collisions may be comparable to the hydrogen-hydrogen collision for higher ionization fractions ($x_{HII} \gtrsim 10^{-2}$), in such circumstances the overall collisional contribution is negligible in comparison to the $Ly\alpha$ pumping effect \cite{smith1966}.

When the first stars are formed, they produce a large amount of the $Ly\alpha$ flux. Due to the WF effect, by absorbing the $Ly\alpha$ photons, hydrogen goes from the ground state to the excited state and returns to the ground state with the hyper-fine effect and eventually emits a 21-cm photon. In redshifts between 12 and 20, the spin temperature is equal to the kinetic temperature, and $x_{\alpha}$ has a non-zero value. For other redshifts, the relation between the spin temperature, the kinetic temperature, and the color temperature is $T_{C}=(1+2.5T_{k})T_{S}/(1+2.5T_{S})$ \cite{Liu:2022iyy}.

The $ly\alpha$ coupling coefficient can be written as follows
\begin{equation}
\label{e8}
x_{\alpha}=\frac{16\pi^{2}T_{*}e^{2}f_{\alpha}}{27A_{10}T_{\gamma}m_{e}c}S_{\alpha}J_{\alpha},
\end{equation}
where $J_{\alpha}$ is the specific flux evaluated at the $Ly\alpha$ frequency and $S_{\alpha}=\int dx\phi_{\alpha}(x)J_{\nu}(x)/J_{\infty}$ is a correction factor to order unity that describes the precise spectrum form in the vicinity of the resonance, or in better words the WF effective coupling \cite{Chen:2003gc,Hirata:2005mz}. We have explaied this parameter in \ref{SEC4}. $J_{\infty}$ is the flux away from the absorption feature, $f_\alpha$ represents the oscillator strength of the $Ly\alpha$ transition which is  0.4162 \cite{Lopez-Honorez:2018ipk}, $m_e$ is the electron mass, and $c$ is the speed of light.

\section{Epoch of Reionization \label{SEC3}}
The formation of galaxies in the early universe is intricately linked with the EoR when gas in the IGM that had been neutral since the recombination at $z \sim 1100$ became once more ionized. After the recombination, the Dark Ages of the universe commence. Roughly, five hundred billion years pass before the first stars and galaxies begin brightening the universe. During the Dark Ages, the dark matter particles form gravitational potential wells by collapsing into halos. As the mass of the dark matter halos increases, only the thermal pressure prevents the gas from collapsing into the halos as well.

For the gas to begin collapsing, the mass of the dark matter halos must reach a threshold at which gravity can surpass the thermal pressure. This mass limit, where the thermal pressure of the gas is exactly compensated by gravity, is known as the "Jeans mass" \cite{Moore:1993sv}. Hydrogen and helium atoms formed during primordial nucleosynthesis made up the majority of the first stars. As they are the first to produce heavier elements through nucleosynthesis, the gas from which they are formed is metal-free and cools less efficiently than gas-containing metals, resulting in the formation of stars that are more massive than stars that exist today \cite{Abel:2001pr}.

The dark matter halos gradually become heavier over time, allowing the formation of not only stars, but also larger structures. While the stars coalesce into clusters, the resulting protogalaxies cool their atoms effectively. Supernova explosions within their origins often decrease the amount of gas in the future galaxy-hosting halos, delaying the start of the star formation process \cite{Kimm:2016kkj}.
It is the transition between the ground state and the lowest excited state of the hydrogen atom that produces the 121.5 nm Lyman-line. Only the neutral hydrogen is capable of this transition. Hence, the presence of the neutral hydrogen clouds along the line of sight is indicated by absorption or emission characteristics at the $\lambda_\alpha$ wavelength. 

Between 0.10 and 0.20 Gyr after the Big Bang ($30 \gtrsim z \gtrsim 20$), the EoR commences, as the first sources emit radiation begin to ionize their surroundings. Halos emit photons that cause ionizing the IGM, leading to $n_{HI} \rightarrow 0$. During this time, the ionization fraction, $x_{HII}$, rather than the spin temperature dominates $\delta T_b$ fluctuations. At the point of termination of the EoR, $n_{HI}\sim 0$ eliminates the 21-cm signal \cite{Doussot:2021kew}. Of course, it will not be exactly zero so 21-cm spectra exist during this time.

In this way, the ionization process takes the form of a mostly spherical ionized area, also known as an ionization bubble \cite{Chuzhoy:2005wv}, which expands outward from the source, but maintains a sharp ionization front beyond which makes the IGM extremely neutral. There are small neutral clumps in the IGM with a considerably higher number density by extension, substantially greater recombination rate; these are called Lyman limit systems \cite{Mittal:2020kjs}.  
 Being small, neutral islands within the ionization bubbles, these can reduce the photons mean free path, thus, slowing down the ionization process \cite{McQuinn:2011aa}.
 
 On a larger scale, photo-ionization and recombination dominate the evolution of the ionized region surrounding a source. When the recombination in the ionized region balances the photon production rate, equilibrium is reached. Accordingly, the volume of the ionized region known as the $Str\Ddot{o} mgren$ sphere is determined.
Although ionizing photons can theoretically originate from a variety of astrophysical sources, the currently accepted theory holds that galaxies provide the vast majority of these photons 
 \cite{finkelstein2016,finkelstein2015c}. Therefore, the formation and evolution of the first galaxies in the universe can be greatly informed by learning about the spatial nature and temporal history of the reionization.

The CMB studies will never be able to detect the reionization directly by imaging individual HII regions. On the other hand, measurements of the CMB anisotropies can enhance constraints on the global features of the reionization. A more accurate measurement of the CMB polarization anisotropies at low multiples possesses the potential to provide some information on the time period of the reionization in addition to stricter constraints on the average redshift of the reionization \cite{Keating:2005ds}. The optical depth, as constrained by the CMB at the present time, is compatible with an instantaneous the reionization redshift of $7.11_{-0.75}^{+0.91}$ \cite{Planck:2018nkj}.

Simulations demonstrate that the reionization possibly begins when overdense regions first formed large HII regions ($z>10$) and ended when the ultimate neutral IGM relics were ionized ($z \sim 6$) \cite{Finlator:2009tr,Alvarez:2008qu}. In the process of ionizing, the reionization probably inhibits star formation in the smallest halos. Those halos incapable of protecting themselves against the abrupt and intense UV background would have all of their gas heated, making them incapable of continuing to form stars. This has important implications for the faint end of the high-redshift luminosity function, which must terminate at some points \cite{oshea2015}.

There are numerous observational probes of the EoR, and this paper focuses on some of them.
It is well established that quasars provide a continuous signal at wavelengths of the same order of magnitude as $\lambda_\alpha$. This variable shows the wavelength of the $Ly\alpha$ line.
A prominent absorption blueward pattern of the $Ly\alpha$ (but redward of the $Ly\beta$) can be seen from rest-frame wavelength, which becomes increasingly crucial when studying the flux emanating from high redshift ($z>6$) quasars \cite{McGreer:2014qwa}. As a result of quasar radiation, the neutral hydrogen is more ionized than the IGM, making observation of the EoR more feasible. This feature is called the "Quasars Gunn-Peterson effect" \cite{Gunn:1965hd}. This effect has made it possible to constrain $x_{HII}=0.6$ at $z = 7.08$, for illustration \cite{Greig:2016vpu}.

Multiple absorption characteristics in the spectra of quasars observable beyond the Lyman line, indicating the presence of the neutral clouds in the IGM along the line of sight, are described by the term "$Ly\alpha$ forest". These absorption lines increase in number with the redshift, converging at $z \gtrsim 6$ into the Gunn-Peterson trough. Subsequently, the Lyman-forest is the "post-reionization version" of the Gunn-Peterson trough, it cannot be used to directly constrain the EoR, although it may be used to provide a lower bound on the patchy remaining neutral fraction after the EoR, with a value of $x_{HII} > 0.94$ at $z = 5.9$ \cite{Greig:2016wjs}.

Photons emitted from the last scattering surface have spanned the EoR in order to reach the observer. During their journey, these photons have a ten percent probability of scattering off free electrons in the IGM at the mean density. This scattering called "Thompson scattering", causes a polarization of the observed CMB at large angular scales. Based on the Plank 2018 measurement $\tau = 0.054\pm 0.007$  \cite{Planck:2018nkj}, one can deduce that the EoR contributes 0.014 to the overall optical depth.

The hyper-fine 21-cm line of HI is one promising source for finding out about the reionization history. HI is present in appreciable amounts throughout most of the reionization era. In addition, because 21-cm radiation is a spectral line, frequency information immediately provides redshift, and radio interferometry may make 21-cm inhomogeneities observable across a range of angular scales. To investigate the IGM at high redshifts, many have turned to the 21-cm line \cite{wyithe2004}.

Furthermore, Gamma-ray bursts (GRBs) \cite{McQuinn:2007gm} and fast radio burst \cite{Heimersheim:2021meu} have been utilized to constrain the EoR. Theoretically, the EoR can be constrained utilizing the Deuterium line \cite{Stancil1998}, Carbon ion $C_{II}$ \cite{Padmanabhan:2021tjr}, $\ce{^3He^+}$ transition \cite{Furlanetto:2006jb}, Hydrogen Deuteride (HD) \cite{Breysse:2021utr}, and neutral oxygen \cite{Doughty:2019pff}.

\section{Wouthuysen-Field Effect\label{SEC4}}

 As previously stated, the $Ly\alpha$ color temperature is predicted to be extremely close to equilibrium with the kinetic temperature of the gas at low redshift, and so the WF effect drives the spin temperature towards the gas temperature in an indirect manner. 
 
 The spin temperature is pushed towards the gas temperature due to collisions between hydrogen atoms. By interacting with the $Ly\alpha$ photons, the WF effect modifies the spin temperature. The inelastic resonant scattering of the $Ly\alpha$ photons, in which the final hyper-fine state is distinct from the initial one, is made possible by electric dipole selection rules. The WF effect strengthens the coupling between the gas and the spin temperatures. Nevertheless, the color temperature diverges slightly from the gas temperature at extremely high redshifts, which modifies the predicted value of the spin temperature \cite{Breysse:2018slj}.

The WF effective coupling is introduced in the following
 \begin{equation}
 \label{e9}
S_\alpha=\tilde{S}_\alpha \frac{\left(T_{C}^{\mathrm{eff}}\right)^{-1}-T_{S}^{-1}}{T_{{K}}^{-1}-T_{S}^{-1}},
\end{equation}

where $\tilde{S}_\alpha$ is evaluated in $10^5 \leqslant \tau_{\mathrm{GP}} \leqslant 10^7$ for continuum photons whcih is described as

\begin{equation}
\label{e10}
\begin{aligned}
\tilde{S}_\alpha= & \left(1-0.0631789 T_{K}^{-1}+0.115995 T_{K}^{-2}\right. \\
& \left.-0.401403 T_{S}^{-1} T_{K}^{-1}+0.336463 T_{S}^{-1} T_{K}^{-2}\right) \\
& \times\left(1+2.98394 \xi+1.53583 \xi^2+3.85289 \xi^3\right)^{-1},
\end{aligned}
\end{equation}

where $\xi=\left(10^{-7} \tau_{\mathrm{GP}}\right)^{1 / 3} T_{K}^{-2 / 3}$ and $\left(T_{C}^{\mathrm{eff}}\right)^{-1}=T_{K}^{-1}+0.405535 T_{K}^{-1}\left(T_{S}^{-1}-T_{K}^{-1}\right)$. $(T_{C}^{\mathrm{eff}})^{-1}$ is functions of $\tau_{GP}$, the spin temperature, and the kinetic temperature. The results from the $(T_{C}^{\mathrm{eff}})^{-1}$ are accurately reproduced here, with one percent accuracy in the Fokker-Planck equation. The last equation is the best-fit equation plotted for the WF effective coupling in terms of the kinetic temperature \cite{Hirata:2005mz}.

The impact of an increase in the neutral hydrogen content of the IGM on the transmission of Lyman photons from galaxies has been the subject of numerous theoretical studies  \cite{McQuinn:2007dy,Santos:2003pc,Furlanetto:2005ir,Mesinger:2007jr}. Studies have shown that when a certain amount of ionizing sources are contained, $Ly\alpha$ can be detected, even in galaxies located in a neutral IGM. Alternatively, the $Ly\alpha$ emission may show a significant recession velocity compared to the absorbing gas \cite{Haiman:2004ap}.

As additional sources were activated at lower redshifts, the early sources in the universe consistently reinforced the ionization exclusion effect.
It is possible that during the EoR, the HII regions could have larger sizes than predicted by the basic model. This could be due to the clustering of sources with high redshift \cite{Sokasian:2003gf}.
Initial projections regarding the influence of the reionization on the luminosity function and line profile of the $Ly\alpha$ emitters were predicated on the supposition that each emitter is situated within its own HII region until the ultimate seepage stage of the final bubble, at which juncture all HII regions combine expeditiously, ending in the completion of the reionization \cite{McQuinn:2006et}.

In the pre-reionized universe, the HII regions in the vicinity of galaxies were not in a state of equilibrium. These regions began to expand when star formation was initiated, and their expansion rate is determined by the ionizing ability of photons. This phenomenon ionizes the neutral IGM more than balancing the recombinations within the ionized region. 

It is anticipated that the HII region will possess a distinct boundary, given the limited mean free path of ionizing photons within the neutral IGM. After the reionization process, the ionizing background significantly ionizes the entirety of the IGM. Moreover, the ionizing flux emanating directly from the galaxy contributes to a proximity effect that affects the ionization balance in its immediate vicinity.

The primary factor influencing the optical depth experienced by the $Ly\alpha$ photons in a galaxy are the size of the surrounding HII region. According to these photons traverse the ionized gas, they undergo minimal absorption and instead experience a redshift ultimately, arriving at the neutral gas in the line wings. The precise amount of absorption is reliant on the size distribution of ionized bubbles throughout the process of reionization. The current analyses adopt an approach that considers each galaxy independently, resulting in relatively small HII regions even during the latter stages of the reionization  \cite{Santos:2003pc}. Nonetheless, both analytical models \cite{Barkana:2003qk} and simulations \cite{Ciardi:2003tp} indicate that galaxies exhibit a high degree of clustering during periods of high redshift.

Existing telescope surveys will also constrain the neutral fraction at higher redshifts, and the James Webb Space Telescope (JWST) will be able to image HII regions undergoing the reionization. Several surveys have focused on the $Ly\alpha$ emitters at epochs when the universe may have been predominantly neutral. These surveys utilize the fact that a galaxy can generate a significant fraction of its flux in the Lyman line \cite{peeples1967}. If the surrounding area of an emitter is predominantly neutral, the transmission of the Lyman line from the galaxy is reduced, and this modification can be used to probe the EoR \cite{Malhotra:2005qf}.

\begin{table}
\begin{center}
\begin{tabular}{ |c|c| }
 \hline
 \multicolumn{2}{|c|}{Initial Condition} \\
 \hline
 \hspace{1cm}Parameter\hspace{1cm} & \hspace{2cm}Value \hspace{2cm} \\
 \hline
 \texttt{(a) BOX\_LEN}  & 100 Mpc    \\
 \texttt{(b) F\_ESC10}&   -1.222   \\
 \texttt{(c) F\_ES7\_MINI} &2.222 \\
 \texttt{(d) ALPHA\_STAR}   &0.5 \\
 \texttt{(e) F\_STAR10} &   -1.25 \\
 \texttt{(f) L\_X} & $10^{40.5} erg.s^{-1}$   \\
 \texttt{(g) L\_X\_MINI} & $10^{40.5} erg.s^{-1}$\\
 \texttt{(h) NU\_X\_THRESH} & 0 \\
 \texttt{(i) Y\_He} & 0.245 \\
 \texttt{(j) OMr} & $8.6\times 10^{-5}$ \\
 \texttt{(k) OMl} & 1-OMb \\
 \texttt{(l) OMb} & 0.04897 \\
 \texttt{(m) hlittle} & 0.6766 \\
 \texttt{(n) H0} & $\texttt{hlittle}\times 3.2407\times 10^{-18} s^{-1}$ \\
 \texttt{(o) m\_p} & $1.67 \times 10^{-24} g$ \\
 \hline
\end{tabular}
\caption{In this table, the initial conditions of our simulation have been shown: (a) is the length of the coeval, (b) is escape fraction, (c) is the escape fraction for minihalos, (d) is the power-law index of fraction of galactic gas, (e) is the fraction of galactic gas in stars for $10^{10}$ solar mass halos, (f) is the specific X-ray luminosity per unit star formation escaping host galaxies, (g) is the specific X-ray luminosity per unit star formation escaping from host galaxies for minihalos, (h) is x-ray energy threshold for self-absorption by host galaxies, (i) is helium fraction, (j) is the relative density of radiation, (k) is dark energy density, (l) is baryon density, (m) is $H_0$/100, (n) is the Hubble parameter, and (o) is the proton mass.}
\label{Table:1}
\end{center}
\end{table}

Molecular hydrogen is destroyed by the LW photons and the required amount of the dark matter is increased for star formation.
The generation of the first stars through the cooling of molecular hydrogen is a highly nonlinear process that can be simulated numerically \cite{Abel:2001pr}. Nevertheless, the potentially lethal effect of the LW background is typically ignored in numerical simulations in which primordial stars are generated. In the special case of a constant $J_{LW}$ intensity, the suppressive effect of the LW background on star formation has been examined \cite{OShea:2007ita}. Minimal cooling mass, $M_{cool}$, the mass of the lightest halo in which stars can form, is increased by the feedback, and the outcomes of these simulations are well represented by the relation

\begin{equation}
\label{e11}
M_{\mathrm{cool}}\left(J_{0}, z\right)=M_{\text {cool }, 0}(z) \times\left[1+6.96\left(4 \pi J_{0}\right)^{0.47}\right],
\end{equation}

where $J_{0}=J_{21}/(10^{-21} erg s^{-1} cm^{-2} Hz^{-1} sr^{-1})$. Besides $M_{\text {cool }, 0}(z)$ represents the minimum cooling mass in the conventional case with no LW background, here $J_{21}$ is the LW intensity \cite{Fialkov:2012su}. This outcome is insufficient for two reasons. One is that it fails to account for the relative velocity, which has a significant effect on the formation of the first stars, among other things, increasing the minimal cooling mass \cite{Stacy:2012iz}. Due to the impact of velocity on star formation, simulated results demonstrate a robust relationship between velocity and the LW flux. The second deficiency of the equation is that it is only valid under the assumption of a constant background intensity during the formation of the halo, whereas in actuality the LW intensity at high redshifts is expected to increase exponentially with time \cite{Visbal:2012aw}. 

Overestimating the strength of the feedback by treating the intensity as fixed at its ultimate value is problematic since the cooling and collapsing associated with the star formation should respond slowly to a decrease in the amount of $H_2$. If the halo core has been cooled and has been collapsed to form a star, for example, modifying the LW flux, may not prevent or reverse this process immediately. Another indicator of the slow process involved is that the cooling time in halo cores is roughly comparable to the Hubble time (which is a pretty large time scale) \cite{Machacek:2000us}.

In the following, we have focused on explaining the simulation of the IGM with the help of the 21cmFAST code. We also elaborate on the initial conditions of the simulation. As well as this, we have mentioned the main effective parameters that play significant roles in the IGM thermal history.

\section{Simulating the IGM\label{SEC5}}

21cmFAST \cite{Murray:2020trn} is an approach for simulating the IGM. It is a semi-numerical code that can generate the 3D cosmological realizations of numerous early universe physical mediums. It is extremely quick and effectively generates density, halos, ionization, spin temperature, 21-cm photons, ionizing flux fields, and so forth by combining the excursion set formalism with perturbation theory. 

It has undergone comprehensive comparison testing with numerical models and has shown excellent agreement at the appropriate scales. 
 The Murchison Widefield Array (MWA) \cite{Tingay:2012ps}, LOFAR \cite{Hothi:2020dgq}, and HERA \cite{HERA:2022wmy} radio-telescopes, among others, have all extensively employed 21cmFAST to imitate the large-scale cosmic 21-cm signal. The speed of 21cmFAST in particular is essential for creating simulations that are large enough (several Gpc across) to simulate contemporary low-frequency observations.

We have demonstrated the initial conditions for our 21cmFast simulation in TABLE \ref{Table:1}. It has been employed for calculations related to the WF effect and then attempted to figure out the correlation between the parameters related to the $Ly\alpha$ and the 21-cm spectrum over time.

This IGM simulation takes several parameters into account, such as the fraction of galactic gas in stars, $f_{*}$, the "escape fraction" or the proportion of ionizing photons that escape into the IGM, $f_{esc}$, and so on which are necessary. If their ionizing photons have high enough escape percentage, early-forming galaxies can be powerful ionizing sources.
The typical (or effective) value of $f_{esc}$ is not yet defined, although it is a major parameter for comprehending the cosmic reionization process \cite{Ouchi:2020yjo}. The fraction of the galactic gas in stars for $10^7$ solar mass minihalos, $f_{*7}$, the specific X-ray luminosity per unit star formation escaping from host galaxies, $L_X$, and the self-shielding factor of molecular hydrogen when experiencing the LW suppression, $f_{sh}$, have been taken into account in the simulation too. These parameters are set by multiplying the reference value by 0, 0.1, 0.5, 1, 2, and 10, and the results have
been compared to each other.
 
It has always been controversial when the EoR began and how long it lasted. Obviously, various models provide different results. Numerous works discover the beginning and the end times of the EoR by analyzing the evolution of the neutral hydrogen fraction. In this work, it has been assumed that the EoR starts when $x_{HI} < 0.95$ and ends when $x_{HI} < 0.1$.

\begin{table}
\begin{center}
\begin{tabular}{ |c|c|c| }
 \hline
 \multicolumn{3}{|c|}{All models} \\
 \hline
 \hspace{0.4cm} Model \hspace{0.4cm} & \hspace{0.1cm} Beginning of the EoR\hspace{0.1cm} & \hspace{0.1cm}Ending of the EoR\hspace{0.1cm} \\
 \hline
$f_{*7\times 0}$ & z=10.42 & z=6.25 \\
$f_{*7\times 1}$ & z=13.21 & z=6.54 \\
$f_{*7\times 0.1}$ & z=10.65 & z=6.39 \\
$f_{*7\times 0.5}$ & z=11.87 & z=6.39 \\
$f_{*7\times 2}$ & z=14.68 & z=6.84 \\
$f_{*7\times 10}$ & z=17.74 & z=8.95 \\
$f_{esc\times 0}$ & z=10.42 & z=6.39 \\
$f_{esc\times 1}$ & z=13.21 & z=6.54 \\
$f_{esc\times 0.1}$ & z=10.65 & z=6.39 \\
$f_{esc\times 0.5}$ & z=11.87 & z=6.39 \\
$f_{esc\times 2}$ & z=14.68 & z=6.84 \\
$f_{esc\times 10}$ & z=18.5 & z=9.55 \\
$L_{X\times 0}$ & z=13.21 & z=6.54 \\
$L_{X\times 1}$ & z=13.21 & z=6.54 \\
$L_{X\times 0.1}$ & z=13.21 & z=6.54 \\
$L_{X\times 0.5}$ & z=13.21 & z=6.54 \\
$L_{X\times 2}$ & z=13.21 & z=6.54 \\
$L_{X\times 10}$ & z=13.49 & z=6.69 \\
$f_{sh\times 0}$ & z=13.78 & z=6.69 \\
$f_{sh\times 1}$ & z=13.21 & z=6.54 \\
$f_{sh\times 0.1}$ & z=13.49 & z=6.69 \\
$f_{sh\times 0.5}$ & z=13.21 & z=6.69 \\
 \hline
\end{tabular}
\caption{In this table, the beginning and the ending redshifts of the EoR are specified for each simulated model gained using the 21cmFAST code.}
\label{Table:2}
\end{center}
\end{table}

As can be seen in TABLE \ref{Table:2}, different models determine significantly various redshifts for the beginning of the EoR, while the end of the EoR typically occurs at redshifts between 6 and 7. This discrepancy indicates the importance of accurately detecting the beginning of the EoR so that the appropriate model can be found in comparison with observations.

It is critical to exclusively focus on this matter; these amounts are derived from TABLE \ref{Table:1} initial conditions. Unmistakably, if these values are altered, new outcomes will emerge. Therefore, improving the precision with these values are determined which is beneficial for achieving improved results.

In this simulation, in addition to contemplating various cosmological parameters, the theory of minihalos has also been utilized. Absorbed light at 21-cm allows us to see collapsed structures in addition to the filaments and sheets that make up the cosmic web. The most required ones are the "minihalos".
Models with the cold dark matter predict a hierarchical structure formation process expressly, low-mass halos collapse first, and the characteristic mass scale growing with cosmic time. Nonetheless, stars are unable to form in the smallest halos because they cannot cool down enough (atomic hydrogen line cooling) without metals \cite{Furlanetto:2002ng}. In the absence of cooling, the gas collapses only to a typical overdensity of 200, leaving star formation impossible.

Likewise the HI clouds in our own galaxy, these minihalos become barely absorbing clouds in the IGM. This condition specifies the upper limit mass for minihalos. The lower limit for minihalo masses is dictated by the Jeans mass, or more accurately the time-averaged Jeans mass, whereas the former is determined in SEC. \ref{SEC3} \cite{Gnedin:1997td}. 

In the absence of the IGM heating, straightforward structure formation models predict that more than 10 percent of the gas should reside in minihalos by $z \simeq 12$. As the IGM is heated, however, the Jeans mass approaches the cooling threshold and minihalos become substantially rarer, for reasons similar to those responsible for the decreased number density of absorption features from the cosmic web \cite{Cen:2002zc}.

 One can anticipate that the numerous neutral hydrogen atoms present in minihalos will generate profound absorption lines, and minihalo absorption is a sensitive probe of the  background radiation through its effect on spin temperature \cite{Villanueva-Domingo:2021cgh}.

Minihalo absorption lines typically have widths of a few kilometers per second (defined by the virial temperature of the minihalo). Minihalos are an independent mechanism for absorption because they are too small to be settled by the majority of numerical simulations. Despite this, semi-numerical simulations involving 21cmFAST code can rectify the issue \cite{Carilli:2004uf}.

In the next section, we have reported the results of the performed simulations, and also, by comparing the evolution of the neutral hydrogen fraction in each model with the data obtained from observation, we have checked the appropriateness of the models.

\begin{figure*}[h!tbp]
  \centering
  \includegraphics[width=15cm]{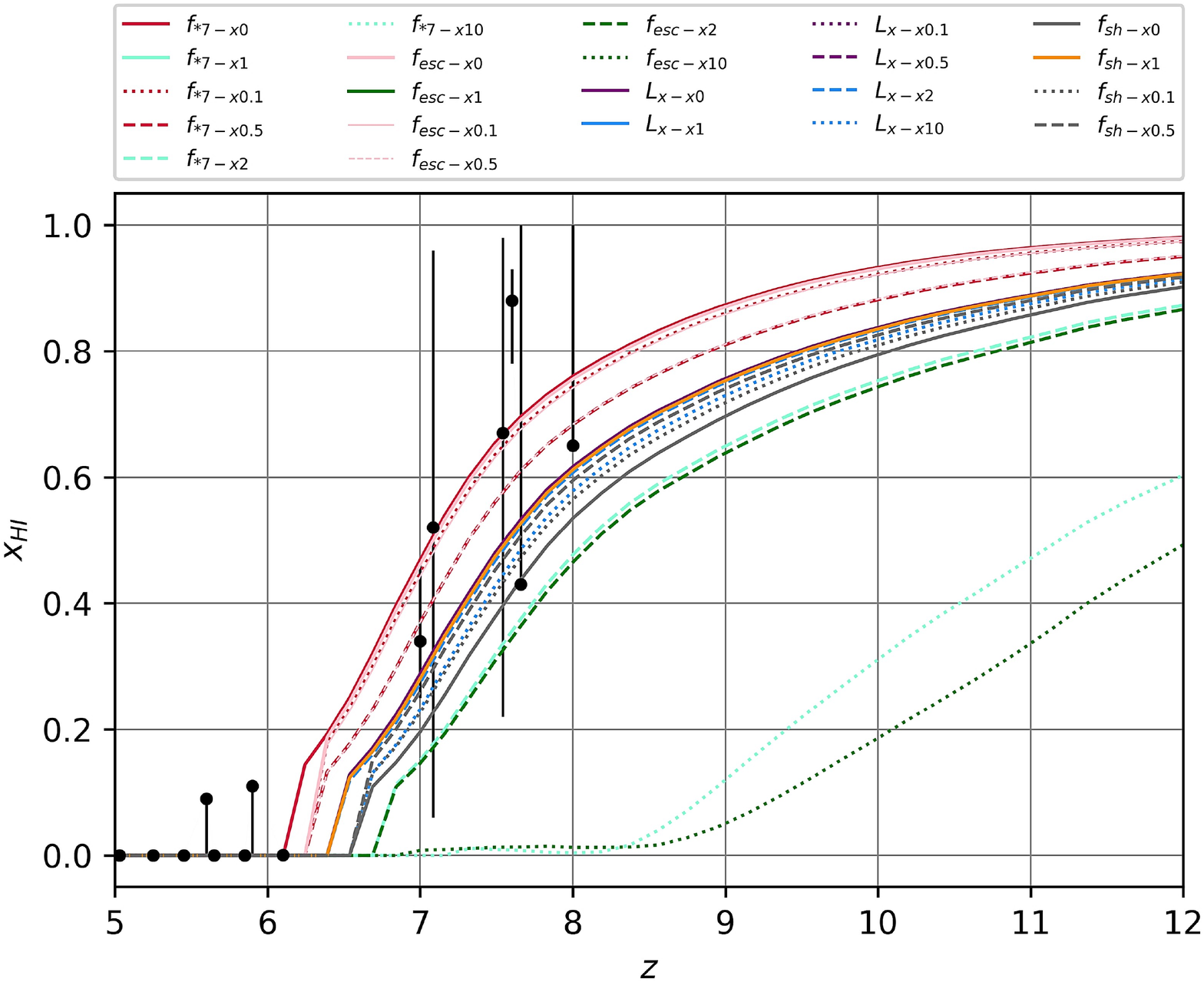}\llap{\includegraphics[height=4.5cm]{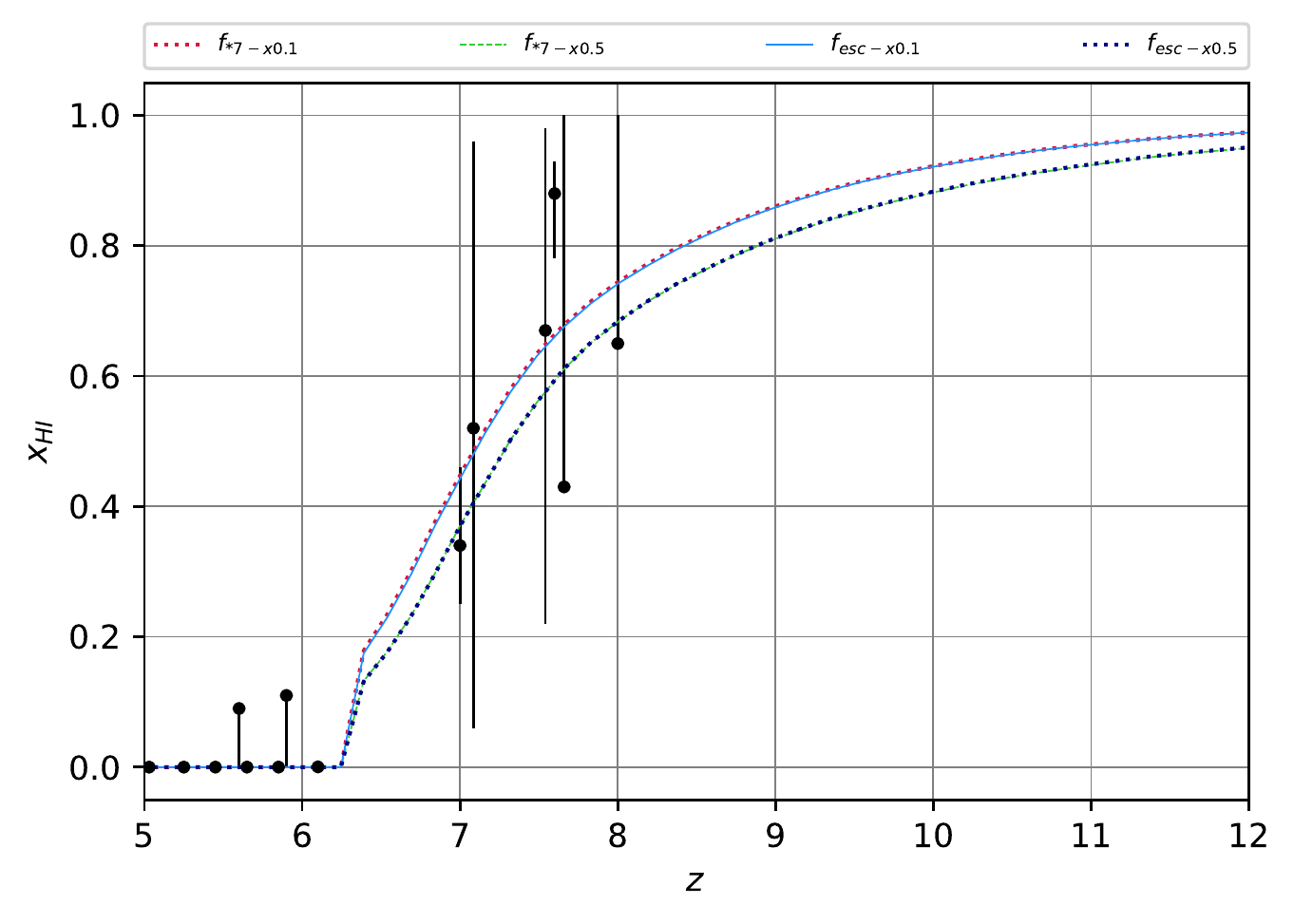}}
   \caption{This figure depicts the trend of the neutral hydrogen fraction variations in the IGM over time for all simulated models. This process of change is also pictured in the sub-figure for the models that are more compatible with the IGM. These four models are $f_{*7\times 0.1}$, $f_{*7\times 0.5}$, $f_{esc\times 0.1}$, and $f_{esc\times 0.5}$. In this figure, observational data are represented as black points with their respective error bars using TABLE \ref{Table:3}.}
   \label{FIG:1}
\end{figure*}

\begin{table}[t!]
\centering
\begin{tabular}{ |c|c|c| }
 \hline
 \multicolumn{3}{|c|}{Observational Data} \\
 \hline
 \hspace{0.5cm} Redshift(z) \hspace{0.5cm} & \hspace{1cm} $x_{HI}$ \hspace{1cm} & \hspace{0.4cm} References \hspace{0.4cm} \\
 \hline
 5.03 & $0.000055_{-0.0000165}^{+0.0000142}$ & \cite{Fan:2005es}\\
 5.25 & $0.000067_{-0.0000244}^{+0.0000207}$ &\cite{Fan:2005es}\\
 5.45 & $0.000066_{-0.0000301}^{+0.0000247}$ & \cite{Fan:2005es}\\
 5.6 &  $0.09_{-0.09}$ & \cite{McGreer:2014qwa}\\
 5.65 & $0.000088_{-0.000046}^{+0.0000365}$ & \cite{Fan:2005es} \\
 5.85 & $0.00013_{-0.000049}^{+0.0000408}$ & \cite{Fan:2005es} \\
 5.9 & $0.11_{-0.11}$ & \cite{McGreer:2014qwa}\\
 6.1 & $0.00043_{-0.0003}^{+0.0003}$ & \cite{Fan:2005es} \\
 7 & $0.34_{-0.09}^{+0.12}$ & \cite{Schenker:2014tda} \\
 7.0851 & $0.52_{-0.46}^{+0.44}$ & \cite{Davies:2018yfp} \\
 7.5413 & $0.67_{-0.45}^{+0.31}$ & \cite{Davies:2018yfp} \\
 7.6 & $0.88_{-0.1}^{+0.05}$ & \cite{Hoag2019} \\
 7.66 & $0.43^{+0.57}$ & \cite{Schenker:2014tda} \\
 8 & $0.65^{+0.35}$ & \cite{Tilvi:2014oia}\\
  \hline
\end{tabular}
\caption{This table shows observational data regarding the neutral hydrogen fraction across different redshifts, as obtained through the techniques explicated in the SEC. \ref{SEC6}. The aforementioned data were utilized to assess the effectiveness of each model.}
\label{Table:3}
\end{table}

\begin{center}
\begin{figure}[h!tbp]
\subfigure[]{
  \includegraphics[width=65mm]{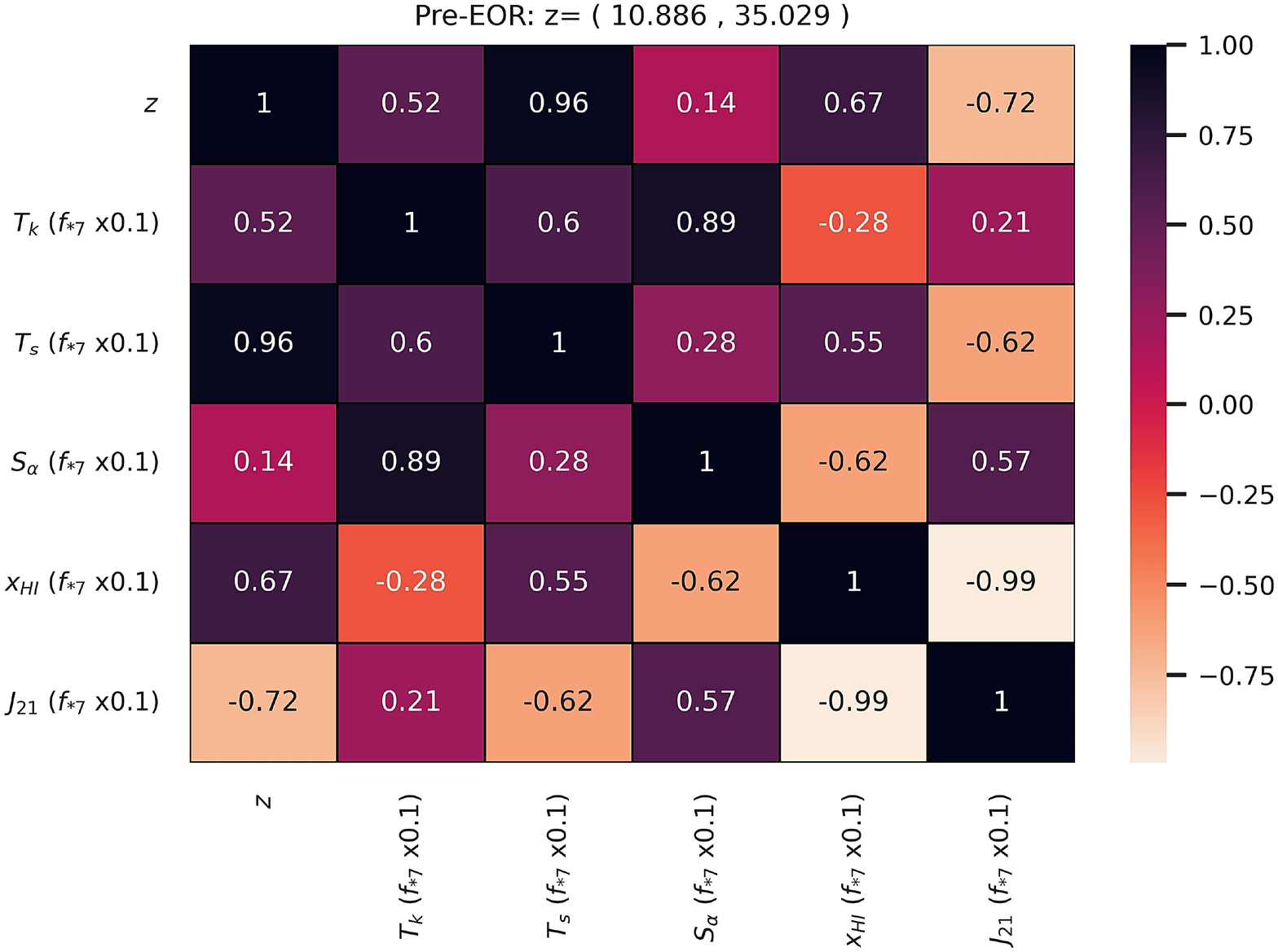}
}
\subfigure[]{
  \includegraphics[width=65mm]{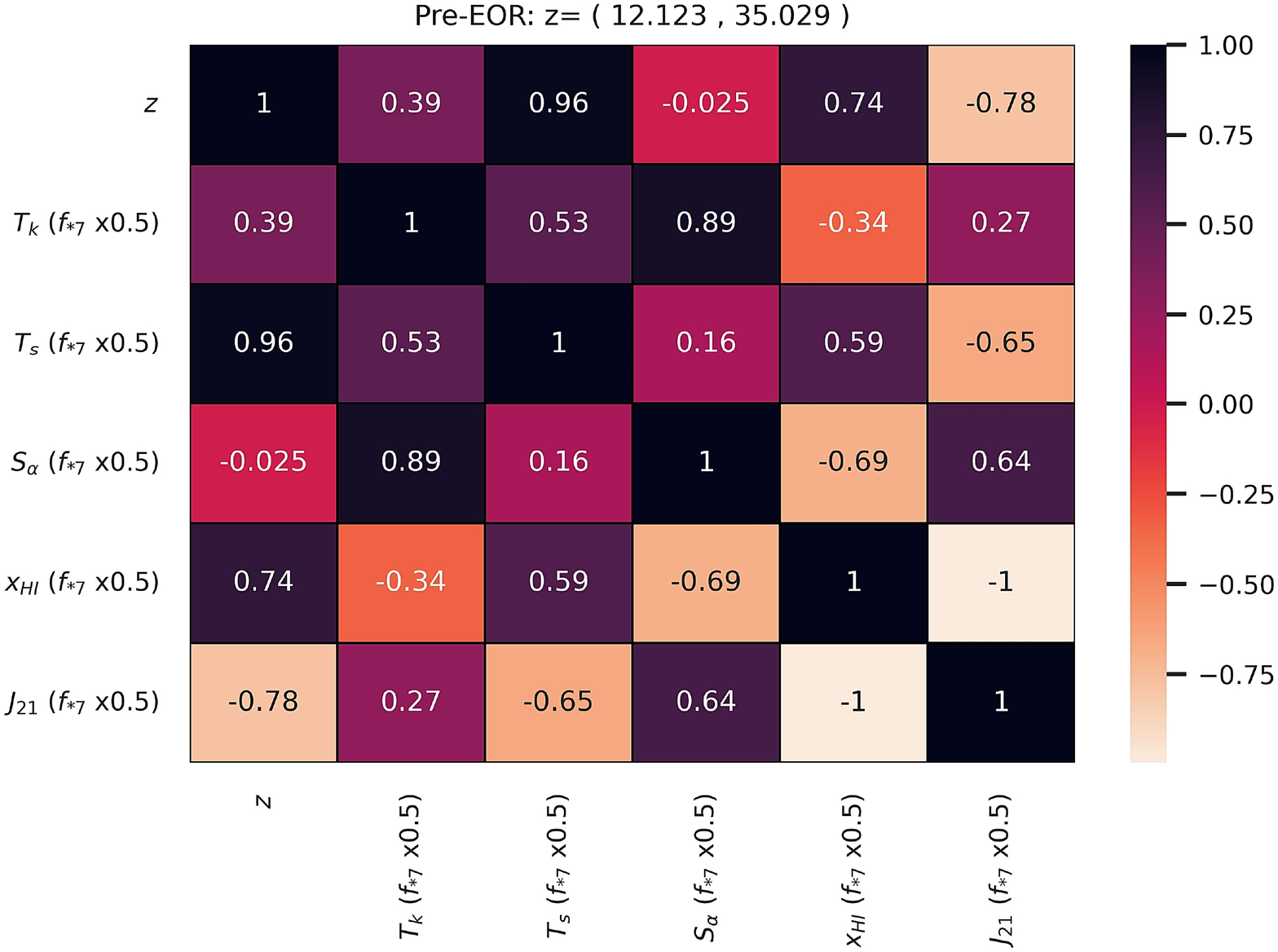}
}
\hspace{0mm}
\subfigure[]{
  \includegraphics[width=65mm]{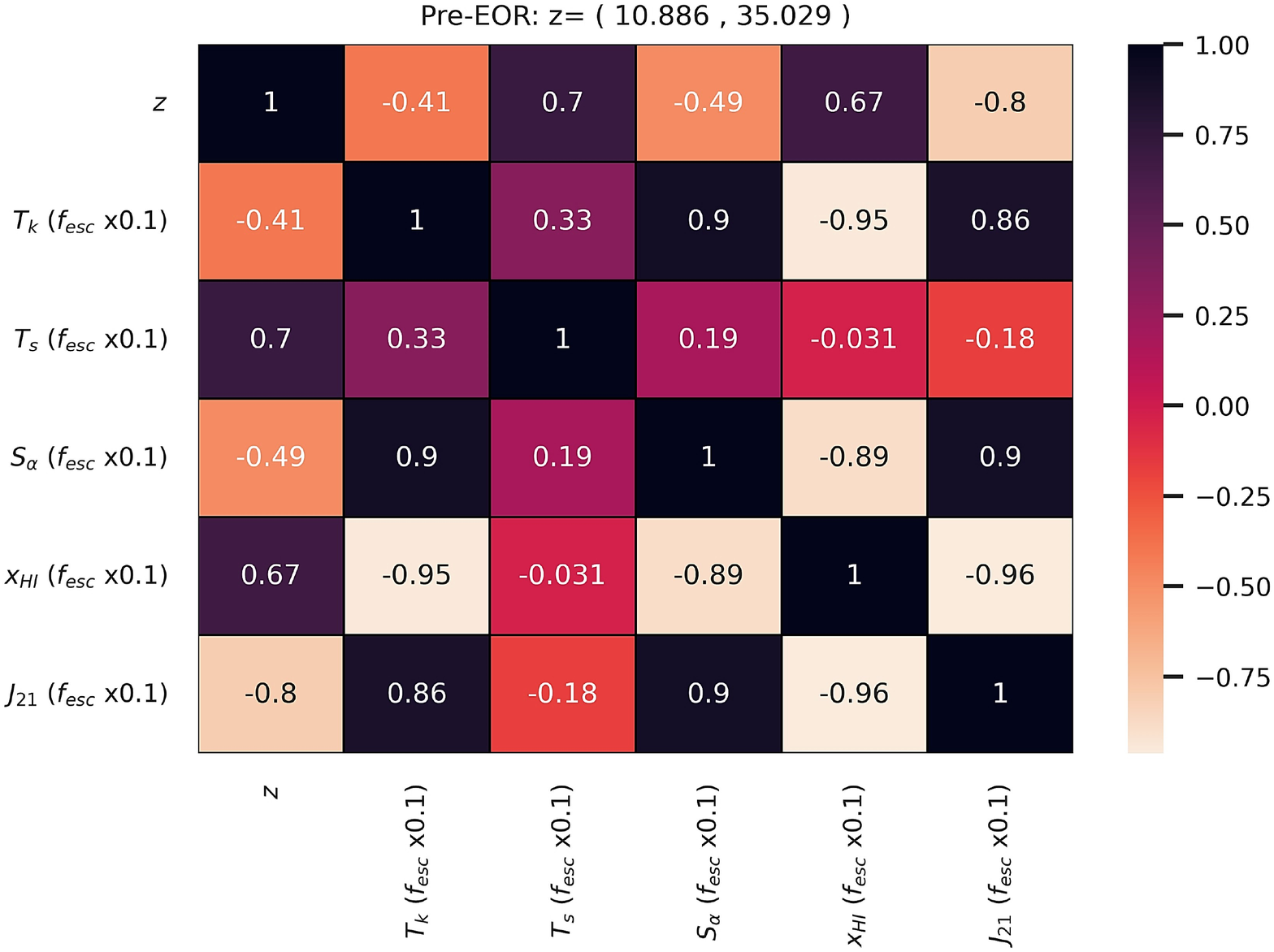}
}
\subfigure[]{
  \includegraphics[width=65mm]{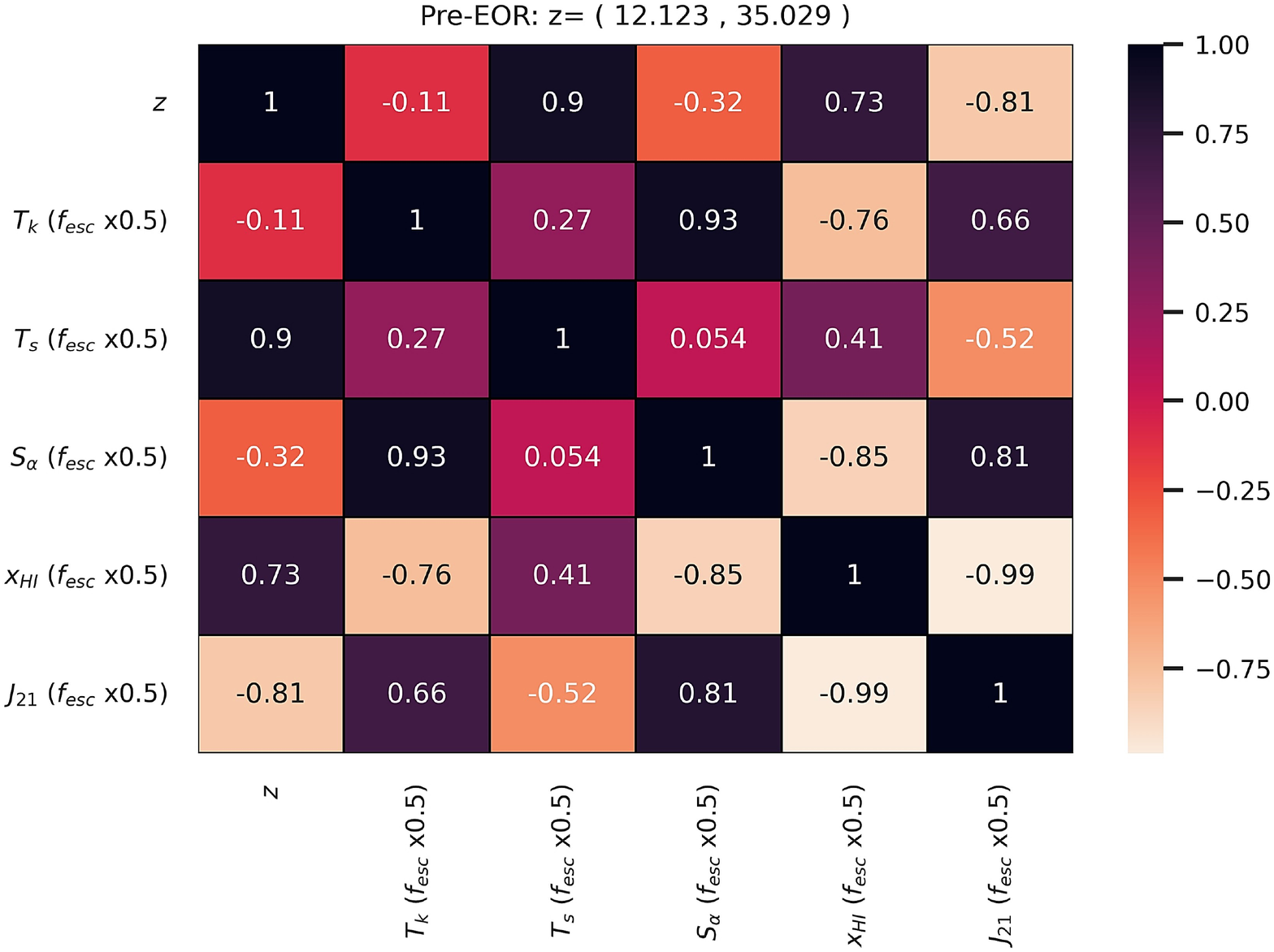}
}
  \vspace*{-3mm}
\begin{center}
\caption{In this figure, correlations between $z$, $T_K$, $T_S$, $S_\alpha$, $x_{HI}$, and $J_{21}$ parameters respectively for 4 models, (a) $f_{*7\times 0.1}$, (b) $f_{*7\times 0.5}$, (c) $f_{esc\times 0.1}$, and $f_{esc\times 0.5}$ are shown in the pre-EoR.}
\label{FIG:2}
 \end{center}
\end{figure}
\end{center}

{\centering
\begin{figure}[h!tbp]

\subfigure[]{
  \includegraphics[width=64.1mm]{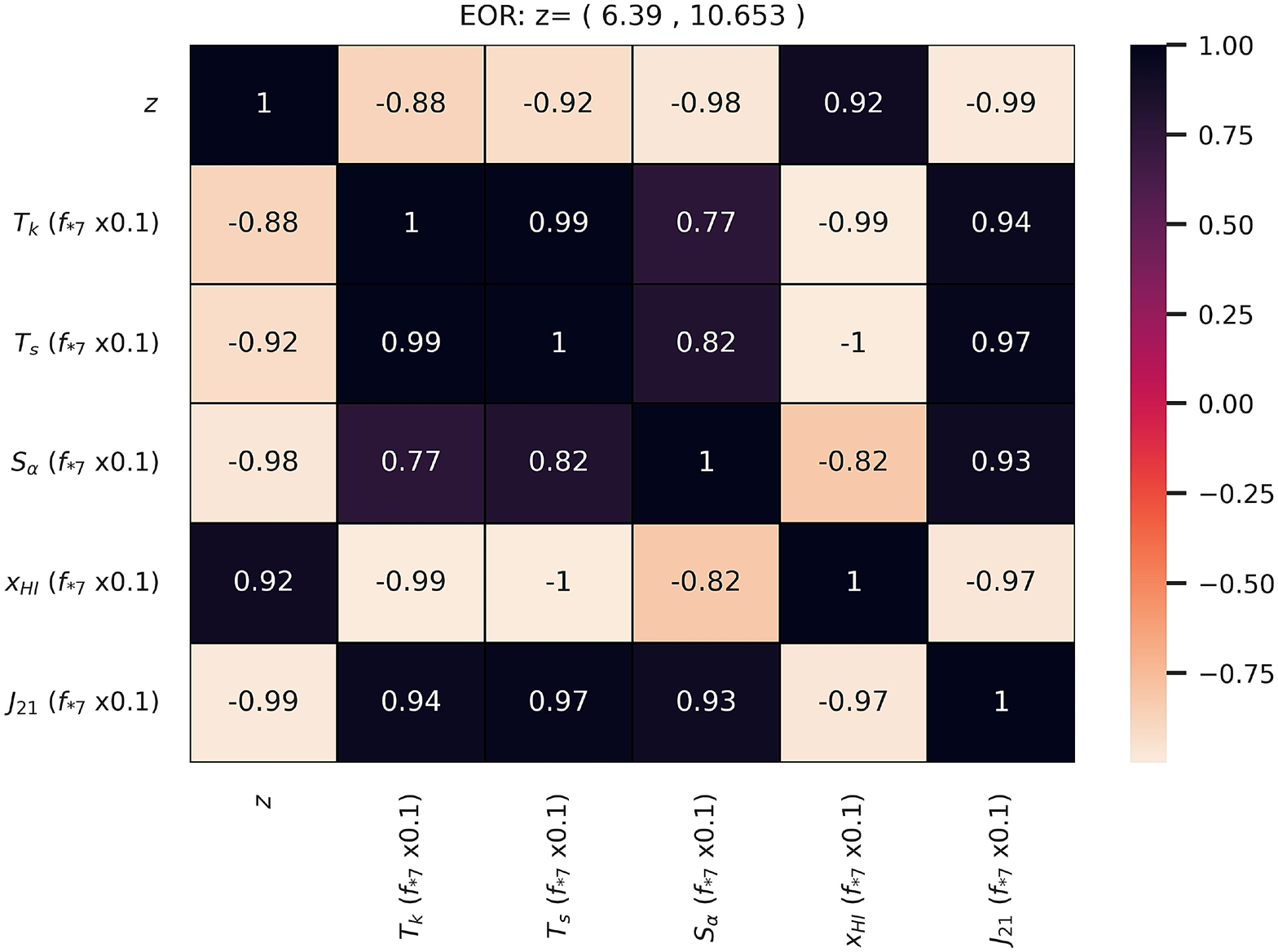}
}
\subfigure[]{
  \includegraphics[width=64.1mm]{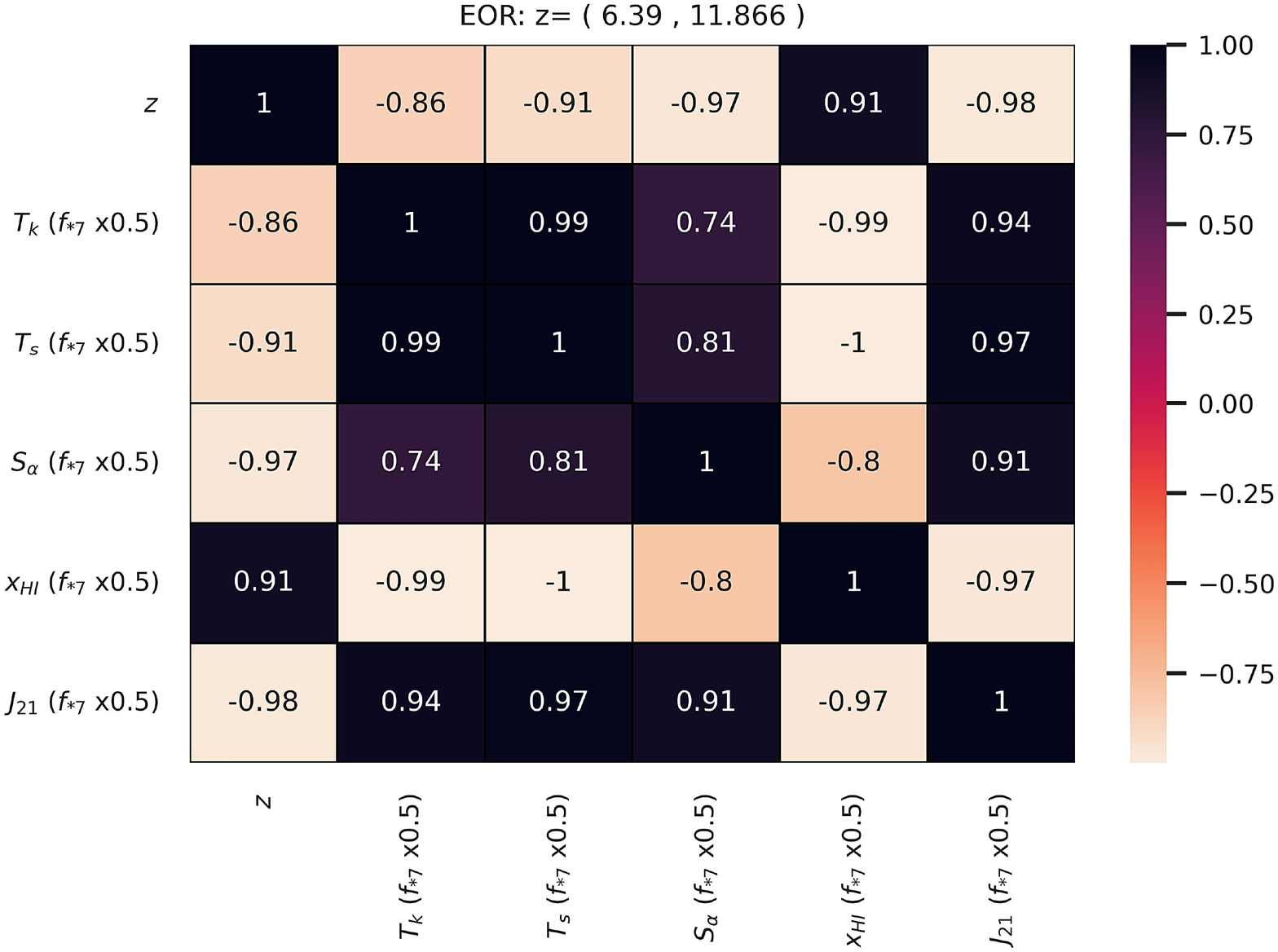}
}
\hspace{0mm}
\subfigure[]{
  \includegraphics[width=64.1mm]{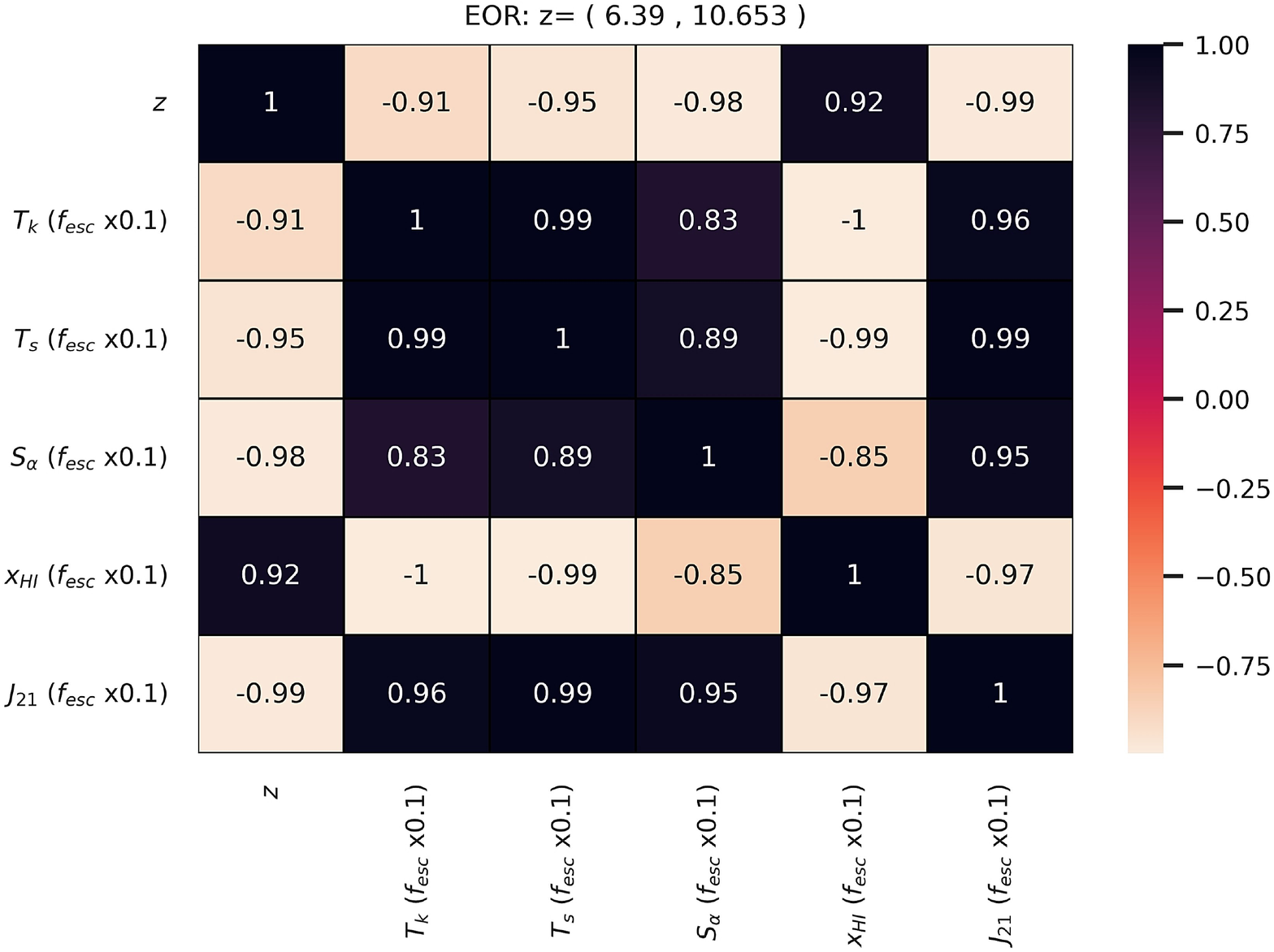}
}
\subfigure[]{
  \includegraphics[width=64.1mm]{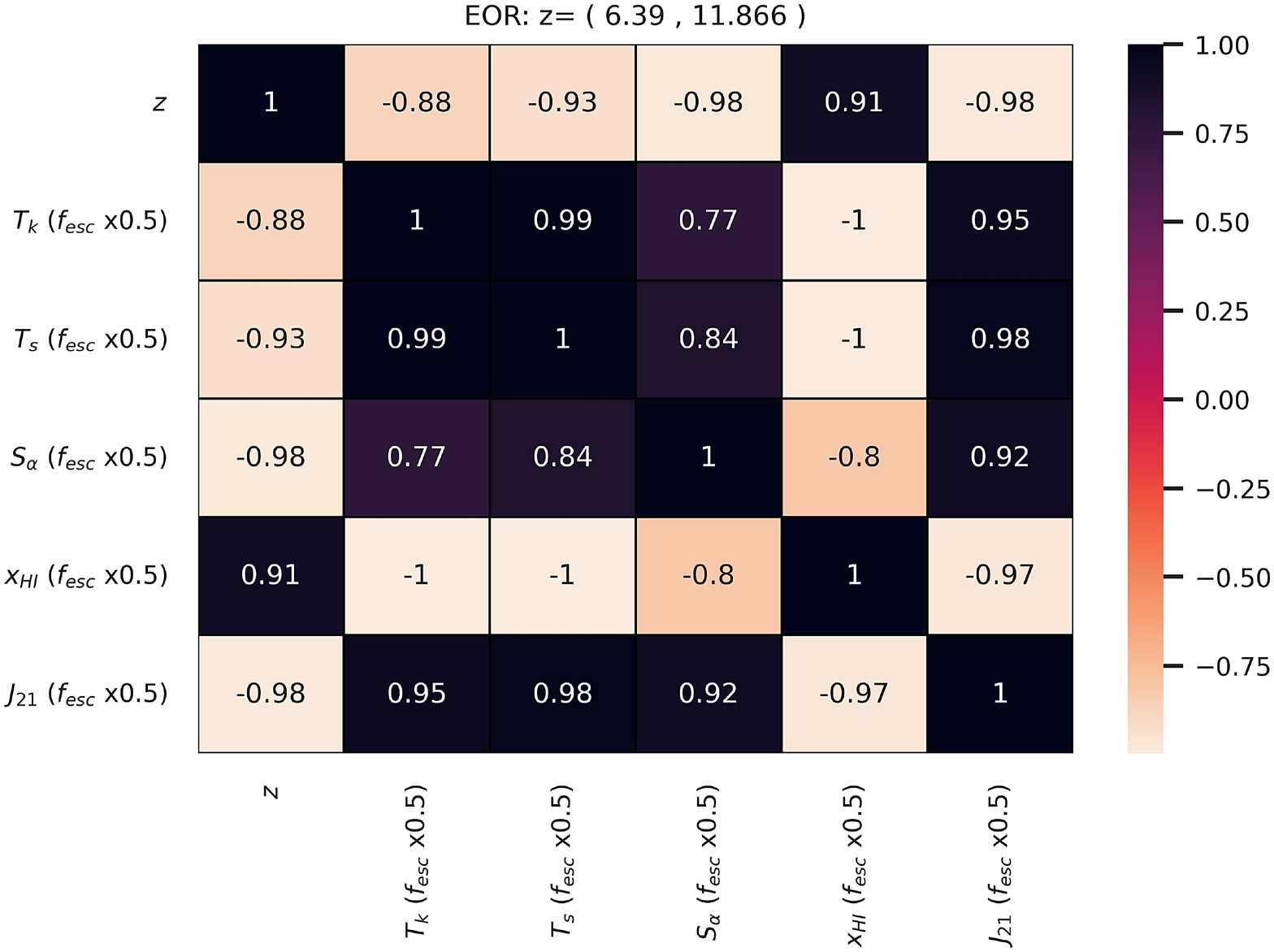}
}
  \vspace*{-3mm}
\begin{center}
\caption{In this figure, correlations between $z$, $T_K$, $T_S$, $S_\alpha$, $x_{HI}$, and $J_{21}$ parameters respectively for 4 models, (a) $f_{*7\times 0.1}$, (b) $f_{*7\times 0.5}$, (c) $f_{esc\times 0.1}$, and $f_{esc\times 0.5}$ are shown in the EoR.}
 \label{FIG:3}
 \end{center}
\end{figure}}

\section{Results\label{SEC6}}
Observational data must be utilized due to the fact that simulations yield radically diverse outcomes.  In the simulation that has been done in this work, the neutral hydrogen fraction can be used as the primary option for comparing results with the observations. The observational data compared to the simulations is indicated in TABLE \ref{Table:3}.
The technique has been used in \cite{Fan:2005es} is considering the Gunn–Peterson optical depth ($\tau = 0.066 \pm 0.013$), in \cite{McGreer:2014qwa} is the dark gaps in quasars spectra, in \cite{Schenker:2014tda,Tilvi:2014oia} is the prevalence of the $Ly\alpha$ emission in galaxies \cite{Bouwens:2015vha}, in \cite{Davies:2018yfp} is quasars continuum, and in \cite{Hoag2019} is the $Ly\alpha $ damping wing.

With both simulation and observational data, we can analyze them and decide which simulated model describes the IGM better. The changes in the neutral hydrogen in all models are displayed in FIG. \ref{FIG:1}. The neutral hydrogen fraction at various redshifts as measured by observations is also shown with error bars. Sub-figure of FIG. \ref{FIG:1} pictures the four models that can explain the IGM history more adequately in comparison with other models. These aforementioned models are $f_{*7\times 0.1}$, $f_{*7\times 0.5}$, $f_{esc\times 0.1}$, and $f_{esc\times 0.5}$.

TABLE \ref{Table:4} highlights the beginning and the ending times of the EoR, given four models yield relatively superior results. As can be seen, the redshift of the end of the EoR is similar in all four models, which is a wildly fascinating result. On the other hand, the starting redshift of the EoR is 10.65 in $f_{*7\times 0.1}$, and $f_{esc\times  0.1}$ models, and 11.87 in $f_{*7\times 0.5}$, and $f_{esc\times 0.5}$ models.

\begin{center}
\begin{table}[h!tbp]
\begin{tabular}{ |c|c|c| }
 \hline
 \multicolumn{3}{|c|}{EoR} \\
 \hline
\hspace{0.5cm} Model\hspace{0.5cm} & \hspace{0.05cm}Beginning of the EoR\hspace{0.05cm} &\hspace{0.1cm} Ending of the EoR\hspace{0.1cm} \\
 \hline
$f_{*7\times 0.1}$ & z=10.65 & z=6.39 \\
$f_{*7\times 0.5}$ & z=11.87 & z=6.39 \\
$f_{esc\times 0.1}$ & z=10.65 & z=6.39 \\
$f_{esc\times 0.5}$ & z=11.87 & z=6.39 \\
 \hline
\end{tabular}
   
\caption{
    This table shows the time of the EoR for chosen models that are more compatible with observational data.}
    \label{Table:4}
\end{table}
\end{center}

To decide on choosing the most descriptive model between these four models, observations at higher redshifts are required though. It is anticipated that in future radio observatories, such as SKA and HERA, it will be possible to reveal the beginning of the EoR with reasonable accuracy.
An accurate cosmological model with known initial conditions benefits greatly from knowing the starting time of the EoR. Taking the stated initial conditions into consideration, it is now possible to deduce which models should be utilized for future work based on the IGM descriptions. Therefore, other models are less likely to produce which may account for the observed IGM.

Now, in a curious endeavor, we are going to investigate the correlation between the neutral hydrogen fraction, the WF effective coupling, the LW flux, the spin temperature, and the kinetic temperature parameters in the pre-EoR FIG. \ref{FIG:2} and the EoR FIG. \ref{FIG:3} for four selected models.

 The goal of this study is to determine the dependency of the neutral hydrogen fraction on the crucial astrophysical parameters of the IGM in different models and at multiple eras. The neutral hydrogen fraction, entirely independent of the model, is completely correlated with the inverse of specific flux evaluated at the $Ly\alpha$ frequency in the pre-EoR.
Furthermore, in all of these models, the neutral hydrogen fraction is strongly correlated with redshift and the WF effective coupling.
In contrast, the neutral hydrogen fraction in the $f_{esc\times 0.1}$, and the $f_{esc\times 0.5}$ models are exceedingly correlated with the inverse of the kinetic temperature. Nevertheless, as $f_{esc}$ increases, this correlation value decreases. This correlation virtually dissipates in two models, $f_{*7\times 0.1}$ and $f_{*7\times 0.5}$.

The correlation between the parameters mentioned in the EoR can be found in FIG. \ref{FIG:3}. This point is very appealing because, regardless of the neutral hydrogen fraction model, it is directly related to redshift and the inverse of other parameters. This demonstrates that the observations associated with each of these parameters during the EoR should correspond with other results derived from the observations of other parameters, and their compatibility should be confirmed to ensure the accuracy and dependability of the results.

What can be a bit worrying is that none of the simulated models touches the $z=7.6$ $x_{HI}=0.88_{-0.1}^{+0.05}$ datum \cite{Hoag2019}. It can be in the most optimistic case that the observation made at that redshift is wrong, and in the most pessimistic case, it is that simulations with other parameters are needed. Consequently, in order to obtain better results, it is essential to consider the observations of the neutral hydrogen fraction, the brightness temperature, and the $Ly\alpha$ photons during the EoR simultaneously.

\section{Conclusion}
The neutral hydrogen fraction is an influential parameter in cosmology and astrophysics. Hydrogen is a suitable representative to analyze the behavior of baryons, as it makes up a large proportion of them. Since baryons are tracers of the dark matter, it is essential to determine the change process of the neutral hydrogen fraction in the formation of structures, and various methods are used to discover its value at varying times. Many observations like determining the Gunn-Peterson optical depth, studying the $Ly\alpha$ emission in galaxies and quasars, etc. are trying to concentrate on investigating the neutral hydrogen fraction in the IGM. 

As well, the beginning and the ending of the EoR can be established with the use of the neutral hydrogen fraction in reconstructing the IGM thermal history. Due to many cosmological models predicting different EoR times, resolving this issue becomes predominant in finding which models are compatible with nature and which cannot accurately describe the universe.

Since the fraction of neutral hydrogen is directly related to the fraction of ionized hydrogen, it is logical to deduce that studies related to ionized hydrogen are also very useful. The HII regions near galaxies in the pre-reionized cosmos were not in a stable equilibrium. The expansion of these zones began with the onset of star formation, and the rate at which they grow is determined not by the equilibrium of the recombinations inside the ionized region, but by the efficiency with which ionizing photons may ionize more the neutral IGM. Due to the short mean free path of ionizing photons in the neutral IGM, it is expected that the HII area will have a clear edge. 

Besides, the local ionization balance is affected by the proximity effect caused by the galaxy's own ionizing flux. The extent of the HII zone encircling a galaxy is the key factor that controls the optical depth experienced by the Lyman photons in that galaxy. The line-winged photons, according to the model, travel through the ionized gas with negligible absorption and arrive at the neutral gas through redshift. 

The spin temperature, the $Ly\alpha$, and the kinetic temperature all play substantial roles in the neutral hydrogen fraction. As can be seen in Eq. \ref{e6}, the spin temperature can change the neutral hydrogen fraction, and the radiation flux that reaches the neutral hydrogen excites them, provide an important point.
Neutral hydrogen fraction depends on the temperature of the CMB photons, the temperature of hyper-fine transition, and the hydrogen-hydrogen de-excitation rate. The importance of this parameter is demonstrated by the fact that changes in any of these factors can affect the variation of 
 the neutral hydrogen fraction during different times. 

Some main future observatories, including PAPER, SKA, and HERA will work to determine the neutral hydrogen fraction. Another method is simulation. One of the greatest codes for determining the neutral hydrogen fraction is 21cmFAST. The initial conditions used for this work can be found in TABLE \ref{Table:1}. In different cases, we have altered the parameters $f_{*7}$, $f_{esc}$, $L_X$, and $f_{sh}$, and TABLE \ref{Table:2} contains a list of all models as well as the redshift of the beginning and the end of the EoR in each model. Multiple observational procedures were utilized to gather the data presented in TABLE \ref{Table:3}. We wished to determine which of the simulated models is most consistent with the observed data. We have compared these models in FIG. \ref{FIG:1}, and the best ones that are more compatible with the IGM thermal history are $f_{*7\times 0.1}$, $f_{*7\times 0.5}$, $f_{esc\times 0.1}$, and $f_{esc\times 0.5}$. These four models are illustrated in the sub-figure of FIG. \ref{FIG:1}, and the beginning and ending of the EoR for each model are specified in TABLE \ref{Table:4}.

In these four models, the ending of the EoR occurs at a redshift of 6.39, indicating that the ending of the EoR can be fixed. The beginning of the EoR is the same in $f_{*7\times 0.1}$ and $f_{esc\times 0.1}$ models, and it is 10.65 and in other models $f_{*7\times 0.5}$, and $f_{esc\times 0.5}$ is 11.87 as well.

Despite this, the issue is that there are still not much data and unfortunately, the error bars are large. In order to improve accuracy and develop better models, the error bars must be reduced. Or, if we can obtain reliable data at redshifts greater than 8, we may be able to obtain a model that precisely reproduces the thermal history of the IGM, and in this way, numerous parameters, notably the star formation rate, the process of formation of structures at the high redshift, and the behavior of the $Ly\alpha$ emission through time may be found for much dark matter halo models.

We also attempted to determine whether redshift, the kinetic temperature, the spin temperature, the WF effective coupling, the neutral hydrogen fraction, and the LW flux in the pre-EoR and the EoR were correlated. All correlations are demonstrated in FIG. \ref{FIG:2} for the pre-EoR and FIG. \ref{FIG:3} for the EoR. In the pre-EoR, the neutral hydrogen fraction, which is completely independent of the model, fully correlates with the LW flux. 

Correspondingly, the neutral hydrogen fraction is closely connected with redshift and the inverse of the WF effective coupling in all of these models. In the $f_{esc\times 0.1}$, and the $f_{esc\times 0.5}$ models, however, the neutral hydrogen fraction is highly correlated with the inverse of the kinetic temperature. Nonetheless, as $f_{esc}$ grows, this correlation value decreases. In two models, $f_{*7\times 0.1}$ and $f_{*7\times 0.5}$, this correlation virtually disappears. It is directly correlated to redshift and the inverse of other parameters in the EoR, regardless of the neutral hydrogen fraction model.
The value of correlation is in the ability to identify how different parameters differ over time and to comprehend how they relate to one another. We can always benefit from this issue from both astrophysical and cosmological perspectives.

\end{document}